\documentclass[preprint,aps]{revtex4}
\usepackage{amsfonts}
\usepackage{hyperref}
\usepackage{amsmath}
\usepackage{graphicx}

\begin{document}

\newcommand{\dd}{\mathrm{d}}
\newcommand{\bd}[1]{\mathbf{#1}}
\newcommand{\Eq}[1]{Eq.\ (\ref{#1})}
\newcommand{\Eqns}[1]{Eqns.\ (\ref{#1})}

\preprint{Obstacle Avoidance}

\title{Obstacle and predator avoidance by a flock}

\author{Nicholas A. Mecholsky}
 \email{nmech@umd.edu}
 \homepage{http://glue.umd.edu/~nmech}
\author{Edward Ott}
\author{Thomas M. Antonsen, Jr.}
\affiliation{
Institute for Research in Electronics \& Applied Physics,\\ University of Maryland, College Park, MD, 20742}
\date{6/17/2009}

\begin{abstract}
The modeling and investigation of the dynamics and configurations of animal groups is a subject of growing attention.  In this paper, we present a continuum model of flocking and use it to investigate the reaction of a flock to an obstacle or an attacking predator.  We show that the flock response is in the form of density disturbances that resemble Mach cones whose configuration is determined by the anisotropic propagation of waves through the flock.  We analytically and numerically test relations that predict the Mach wedge angles, disturbance heights, and wake widths.  We find that these expressions are insensitive to many of the parameters of the model.
\end{abstract}

\pacs{Valid PACS appear here}

\maketitle
\section{Introduction}
The flocking of biological organisms into groups has been a phenomenon of long standing interest.  Birds, fish, bacteria, and certain robots exhibit rich collective behavior.  Research in this area has generally employed two main modeling paradigms:  discrete individuals and continuous densities of individuals \cite{Reynolds87, Vicsek95, Couzin02, Dorsogna06, Chuang07, Bertin06, Levine00, Flierl99, Tanner03, Topaz04, Topaz08, Chen06, Mogilner03, Erdmann02, Ratushnaya06, Ramaswamy02, Eckhardt06}.  While both approaches are useful and contribute to a more complete description of flocking, we focus on a continuum approach.  In this paper, we will consider the response of a flock to a stationary or moving `obstacle'.  In the case of a moving obstacle, our considerations might also be considered as modeling the avoidance response of a flock to a predator.  Past research investigating obstacle avoidance has employed a discrete approach \cite{Reynolds87, Lee06, Lee2006270, Zheng05}.  As compared to a discrete description, the continuum description has the advantage of economically treating very large numbers of individuals and is, in some cases, easier to treat analytically and to interpret.  Its disadvantage is primarily that taking the continuum limit is an abstraction from the real case of discrete flock members.

The current paper introduces a moving obstacle into a large flock and studies the effect of this obstacle on the flow around the obstacle.  We model the obstacle as a localized region exerting a repulsive `pseudo-force' on the flock continuum.  Using our description, we are able to describe the propagation of information in terms of a few parameters in the model.  To do this, we use a fluid characterization of a flock.   For a review of this type of approach, as well as other approaches to modelling flocks, see \cite{TonerReview05}.

To facilitate our analysis we will utilize a linearized theory in which the flock response to the obstacle/predator pseudo-force is assumed to be proportional to the pseudo-force strength.  That is, the obstacle/predator is treated as a linear perturbation.  Results obtained through this type of analysis are expected to yield qualitative insights to the dynamics of the full nonlinear problem, and may also yield quantitative understanding in the region far enough from the obstacle/predator where the perturbations become small.  In the next section, we will introduce our continuum description of the flock.  In the following sections, we explore the small amplitude wave dispersion relation and derive an expression for the disturbances that propagate through the flock.  Next, a linearized response is investigated and the resulting density perturbation is analyzed analytically.  Finally, results of numerical evaluation of the density perturbation are presented and compared to the theory.

\section{Continuum flocking equations}\label{sec:2}
The equations we consider for flocking in three dimensions are
\begin{align}\label{fullsystem1}
\frac{\partial \bd{v}}{\partial t} + \bd{v} \cdot \nabla \bd{v} &= \frac{1}{\tau} \left( 1 - \frac{v^2}{v_0^2} \right) \bd{v} - \frac{1}{\rho} \nabla P(\rho) - \nabla U - \bd{W}(\bd{v}) \\\label{fullsystem2}
\frac{\partial \rho}{\partial t} + \nabla \cdot \left( \rho \bd{v} \right) &= 0,
\end{align}
where $\rho$ is the number density of flocking individuals, $\bd{v}$ is the macroscopic vector velocity field of the flock, $v$ is its magnitude, and $v_0$ and $\tau$ are constants. The basic structure of these partial-differential equations includes terms that define the acceleration of the fluid density of the flock, along with continuity of flock members.  The right-hand side of \Eq{fullsystem1} consists of four `pseudo-forces' representing speed regulation, pressure, pairwise attraction, and a `non-local viscosity' term.  These terms are discussed below. 

The first term on the right-hand side of \Eq{fullsystem1} acts as a speed-regulation term used commonly in the literature \cite{Erdmann03, Erdmann02, Chuang07} and apparently first used by Rayleigh \cite{RayleighSound} as cited by \cite{Erdmann02}.  This term either increases or reduces the magnitude of the velocity depending on how the velocity compares to $v_0$.  If $v>v_0$, the acceleration is negative in the direction of $\bd{v}$, and thus $|\bd{v}|=v$ is reduced.  If $v<v_0$, the acceleration is positive in the direction of $\bd{v}$, hence $v$ is increased.  Thus $v_0$ can be regarded as modeling the average preferred natural speed of an individual.  The time scale for this velocity clamping is $\tau$.  Note that this speed-regulation term is frame dependent and applies when considering the frame in which the medium (e.g., air for birds, water for fish, or land for ungulates), through which the flock individuals move, is stationary.

In order to model the presumed tendency of nearby flock members to repel each other to avoid collision, some past models have introduced a pressure-like term, as in the second term on the right-hand side of \Eq{fullsystem1}.  Examples can be found in \cite{TonerReview05}. In addition, another means to model repulsion is via a general repulsive potential; i.e., a pairwise non-local soft-core potential (see \cite{Mogilner03,Levine00,Dorsogna06}).  We model repulsion using a pressure term, $P(\rho)$.  For future reference, we write the pressure as a Taylor series around a density $\rho_{0}$ as
\begin{equation}
P(\rho) = c_s^2 \delta \rho + \frac{\partial^2 P}{\partial \rho^2} \bigg|_{\rho = \rho_{0}} \delta \rho^2 + \ldots
\end{equation}
where $c_s^2 = \frac{\partial P}{\partial \rho} \big|_{\rho = \rho_{0}}$ and $\delta \rho = \rho - \rho_0$.

The third term on the right-hand side of \Eq{fullsystem1} is a long-range attractive pseudo-force where long range attraction is used to model the tendency for flocks to form.  This force is taken to be due to an attractive pseudo-potential, $U$, which is of the form 
\begin{equation}
U(\bd{x}) = \int u(\bd{x} - \bd{x}') \rho(\bd{x}') d\bd{x}'.
\end{equation}
It proves convenient to choose the kernel $u(\bd{x} - \bd{x}')$ to satisfy the modified Helmholtz equation,
\begin{equation}
\left( \nabla^2 - \kappa_{\rho}^2 \right) u(\bd{x} - \bd{x}') = 4 \pi u_0 \delta \left( \bd{x} - \bd{x}' \right),
\end{equation}
where $u_0>0$ is the strength of the potential.  In three dimensions $u(\bd{x}-\bd{x}')$ has the form of an attractive exponentially-screened Coulomb potential,
\begin{equation}
u(\bd{x} - \bd{x}') = - u_0 \frac{e^{-\kappa_{\rho} |\bd{x} - \bd{x}'|}}{|\bd{x} - \bd{x}'|}.
\end{equation}
The quantity $\kappa_{\rho}^{-1}$ provides a long-distance cutoff to the attractive pseudo-force.  This type of attractive potential has been used in previous continuum flocking models \cite{Mogilner03, Levine00, Dorsogna06}.

Similar to the non-local attractive potential, we model the presumed tendency for nearby flock members to attempt to align their velocities by use of the term 
\begin{equation}
\bd{W}(\bd{x}) = \int w(\bd{x} - \bd{x}') [\bd{v}(\bd{x}') - \bd{v}(\bd{x})] \dd \bd{x}',
\end{equation}
with the kernel $w(\bd{x}-\bd{x}')$ satisfying an equation similar to that for the attractive kernel,
\begin{equation}
\left( \nabla^2 - \kappa_{w}^2 \right) w(\bd{x} - \bd{x}') = 4 \pi w_0 \delta \left( \bd{x} - \bd{x}' \right),
\end{equation}
with strength $w_0 > 0$ and screening length scale $\kappa_{w}^{-1}$.  This term reorients the velocity vector, $\bd{v}(\bd{x})$, toward the average velocity of the other flock members, weighting velocities of flock members closer to $\bd{x}$ more strongly than those farther away.  These are our general equations that model flocking.  Various dynamical behaviors and flocking equilibria can be explored using this framework.  However, in the rest of the paper, we consider perturbations around a specific equilibrium density defined below.

We consider the following simplified situation.  A particular, spatially-homogeneous steady-state solution to Eqs.\ (\ref{fullsystem1}) and (\ref{fullsystem2}) is 
\begin{equation}\label{constsolution}
\rho(\bd{x}) = \rho_0 = \textrm{const}. \quad \textrm{and} \quad \bd{v}(\bd{x}) = \bd{v}_0 = \textrm{const}.
\end{equation}
Alternatively, we may think of this equilibrium as a localized approximation of a more complicated solution where the density is not everywhere constant.  For example, in the middle of a nonuniform flock, the density in equilibrium will be nearly constant (see \cite{Levine00}).  

To the general equations Eqs.\ (\ref{fullsystem1}) and ~(\ref{fullsystem2}), we will add an additional, external, localized, repulsive potential that we view as modeling the effect of a stationary obstacle or a predator moving through the flock with velocity $\bd{v}_p$.  In the next section, we treat this problem within the framework of linearized theory and consider how perturbations propagate through the flock.

\section{Heuristic Discussion of Interaction with an Obstacle}\label{sec:heuristic}
\subsection{Dispersion Relation and Plane Waves}\label{sec:dispersion}
By looking at the dispersion relation of linear waves in the full system, we may determine how such waves propagate within the flock.  This will inform us as to the relationship between the frequency, wavelength, and propagation direction of the waves.  In our case, we will find that this will predict a disturbance cone when an obstacle or predator is encountered by a flock.  For convenience we make a transformation of independent variables such that $\bd{x}' = \bd{x} - \bd{v}_p t$ and $t'=t$ where $\bd{v}_p$ is the velocity of the obstacle or predator relative to the stationary frame in which the preferred flock speed is $\bd{v}_0$.  This is similar to a Galilean frame transformation except that the velocity $\bd{v}$ remains the velocity in the stationary frame.  After making this transformation we drop the primes on $\bd{x}'$ and $t'$.  Writing Eqs.\ (\ref{fullsystem1}) and (\ref{fullsystem2}) in this new frame gives
\begin{align}\label{fullsystemnew1}
\frac{\partial \bd{v}}{\partial t} -\bd{v}_p \cdot \nabla \bd{v}+ \bd{v} \cdot \nabla \bd{v} &= \frac{1}{\tau} \left( 1 - \frac{v^2}{v_0^2} \right) \bd{v} - \frac{1}{\rho} \nabla P(\rho) - \nabla U - \bd{W} \\\label{fullsystemnew2}
\frac{\partial \rho}{\partial t} - \bd{v}_p \cdot \nabla \rho + \nabla \cdot \left( \rho \bd{v} \right) &= 0.
\end{align}
Setting $\rho = \rho_0 + \delta \rho$ and $\bd{v} = \bd{v}_0 + \delta \bd{v}$, with $|\delta \rho| \ll \rho_0$ and $|\delta \bd{v}| \ll \bd{v}_0$, we substitute these into Eqs.\ (\ref{fullsystemnew1}) and (\ref{fullsystemnew2}) and only keep linear terms in $\delta \rho$ and $\delta \bd{v}$.  Taking Fourier transforms in both space and time we obtain
\begin{align}\label{FTedEqnsmoving}
-i \omega \delta \tilde{\bd{v}} + i \bd{k} \cdot \bd{v}_r \, \delta \tilde{\bd{v}} &= -\frac{2}{\tau v_0^2} \left( \bd{v}_0 \cdot \delta \tilde{\bd{v}}\right) \bd{v}_0 - i \bd{k} \left( c_s^2 \frac{\delta \tilde{\rho}}{\rho_0} + \tilde{u} \delta \tilde{\rho} \right) - \nu_w \delta \tilde{\bd{v}}\\\label{deltarhomoving}
0 &= - i \omega \delta \tilde{\rho} + i \bd{k} \cdot \bd{v}_r \, \delta \tilde{\rho} + \rho_0 i \bd{k} \cdot \delta \tilde{\bd{v}},
\end{align}
where we define $\bd{v}_r = \bd{v}_0 - \bd{v}_p$, and $\tilde{f} = \tilde{f}(\bd{k},\omega)$ denotes the Fourier transform of $f(\bd{x},t)$ given by
\begin{equation}
\tilde{f} = \tilde{f}(\bd{k},\omega) = \int f(\bd{x},t) \exp(i \omega t - i \bd{k} \cdot \bd{x}) d\bd{x} dt.
\end{equation}
Hence we have
\begin{align}\label{FTdefs}
\tilde{u} (k^2) &= \frac{- 4 \pi u_0}{k^2 + \kappa_{\rho}^2},\\
\tilde{w} (k^2) &= \frac{- 4 \pi w_0}{k^2 + \kappa_{w}^2},\\
\nu_w (k^2) &= \tilde{w} (k^2) - \tilde{w} (0) = \left( \frac{4 \pi w_0}{\kappa_{w}^2} \right) \frac{k^2}{k^2+\kappa_w^2},\\\label{FTdefsend}
c_s^2 &= \frac{\partial P}{\partial \rho} \bigg|_{\rho = \rho_{0}}.
\end{align}
Using \Eq{deltarhomoving} to eliminate $\delta \tilde{\rho}/\rho_0$ from \Eq{FTedEqnsmoving}, we arrive at
\begin{equation}\label{mastermoving0}
\left[(\omega - \bd{k} \cdot \bd{v}_r) + i \nu_{w} \right] \delta \tilde{\bd{v}} = -\frac{2 i \bd{v}_0 \cdot \delta \tilde{\bd{v}}}{\tau v_0^2} \bd{v}_0 + \frac{(c_s^2 + \rho_0 \tilde{u}) \bd{k} \cdot \delta \tilde{\bd{v}}}{\omega - \bd{k} \cdot \bd{v}_r} \bd{k}.
\end{equation}
Restricting our attention to the case where $\tau \rightarrow 0$ and $\nu_{w} \rightarrow 0$ (the limit in which speed regulation occurs instantaneously and the non-local viscosity is absent), we obtain particularly simple results describing the propagation of waves within the flock.  Note that to accommodate the $\tau \rightarrow 0$ limit in \Eq{mastermoving0}, we must have that $\bd{v}_0 \cdot \delta \tilde{\bd{v}} \rightarrow 0$.  Without loss of generality, we choose $\bd{v}_0$ to be in the $x$ direction, which means that $\delta \tilde{\bd{v}} = \delta \tilde{v}_y \, \hat{\bd{y}} + \delta \tilde{v}_z \, \hat{\bd{z}}$.  If we now project \Eq{mastermoving0} onto the $y$ and $z$ directions, we get two coupled equations for $\delta \tilde{v}_y$ and $\delta \tilde{v}_z$.  These yield the dispersion relation,
\begin{equation}
(\omega - \bd{k} \cdot \bd{v}_r)^2 = k_{\perp}^2 c_s^2,
\end{equation}
where we have taken $\nu_{w} \rightarrow 0$, and defined $\bd{k}_\perp = k_y \hat{\bd{y}} + k_z \hat{\bd{z}}$ giving the magnitude as $k_{\perp}^2 = k_y^2 + k_z^2$.  Also, we have replaced $c_s^2 + \rho_0 \tilde{u} \rightarrow c_s^2$, which is true for large $k$.  We can write the final dispersion relation as
\begin{equation}
\omega = \bd{k} \cdot \bd{v}_r \pm k_{\perp} \, c_s.
\end{equation}
Thus, the group velocity of waves, in the frame moving at a velocity $\bd{v}_p$, within the flock is given by
\begin{equation}\label{dispersion}
\bd{v}_g = \frac{\partial \omega}{\partial \bd{k}} = \bd{v}_r + \frac{\bd{k}_{\perp}}{k_{\perp}} c_s,
\end{equation}
In the next section, we use this result to derive a disturbance cone that propagates through the flock when the flock encounters an obstacle or predator.

\subsection{Mach Cones}\label{sec:machcones}
\begin{figure}[!hb]
\begin{center}
	\includegraphics[width=12cm]{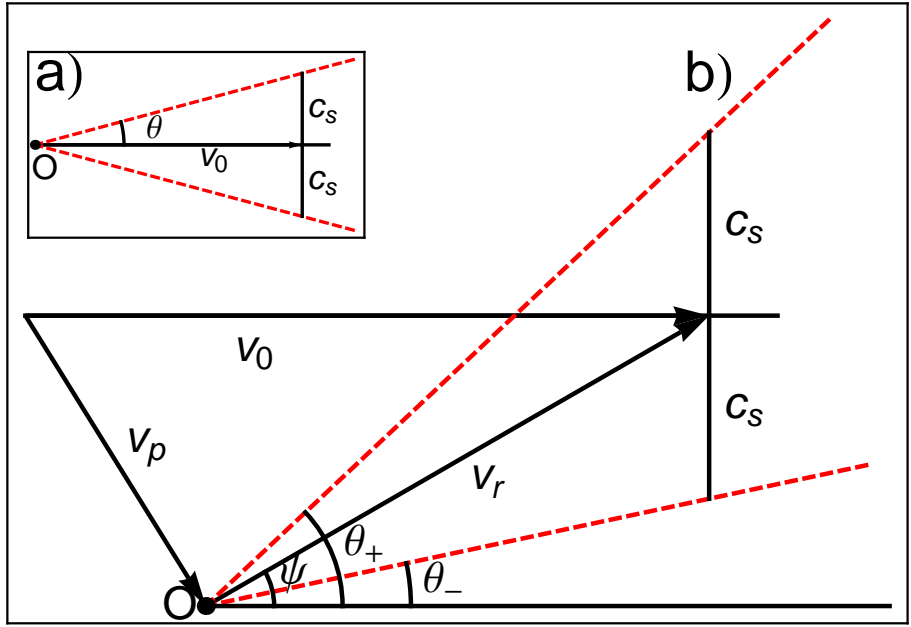}\\
	\caption{\label{fig:conediagram}(color online) Diagram for a density disturbance caused by flock moving past an obstacle in steady state. a) Static obstacle: Obstacle is stationary (at point $O$), the dashed red lines indicate the intersection of a plane passing through the axis of the cone of disturbance and a large-density fluctuation wake that would exist downstream of the flock/obstacle interaction.  The angle $\theta$ is referred to as the `wedge' angle.  b) Moving obstacle: In the frame of the obstacle at point $O$, the density disturbance is indicated by the dashed lines.  $\bd{v}_0$ is the velocity of the flock and $\bd{v}_p$ is the velocity of the obstacle, both in the frame of the medium.}
\end{center}
\end{figure}

Following Mach's well-known construction (see for example \cite{faber}) for the cone produced in supersonic velocities through a fluid, we may develop a prediction for the information cone that is propagating though the flock using the dispersion relation derived above.  In the case of a \emph{stationary} object (Fig.\ \ref{fig:conediagram}(a)), the only way that information can travel is perpendicular to the direction of motion with the propagation speed of $c_s$, in the frame of the obstacle, as can be seen in \Eq{dispersion} with $\bd{v}_r = \bd{v}_0$ (or equivalently $\bd{v}_p = 0$).  Accordingly, we get a right-circular cone in three dimensions (a wedge in two dimensions) of cone angle $\theta$, measured from the direction of the flock, given by
\begin{equation}\label{tantheta}
\tan \theta = \frac{c_s}{v_0},
\end{equation}
where $c_s$ is defined above (\Eq{FTdefsend}).  Notice that this is valid for all velocities, contrary to a usual acoustic Mach cone which only exists for velocities of the moving object that are above the sound speed.  Equation~\ref{tantheta} limits to an angle of $\theta = \frac{\pi}{2}$ for small $v_0$, and $\theta \cong 0$ for large $v_0 \gg c_s$.  

\begin{figure}[!hb]
	\begin{center}
 	\includegraphics[width=12cm]{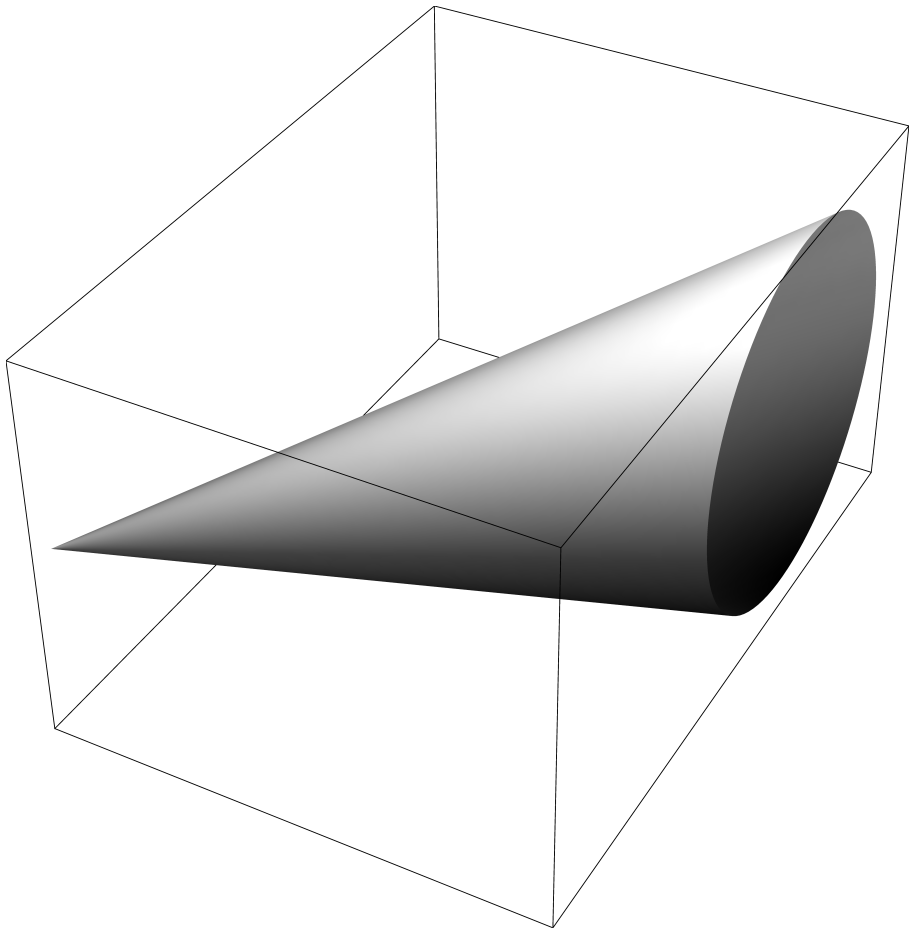}\\
      \caption{\label{fig:oblique}An oblique circular cone}
	\end{center}
\end{figure}

For the case of a moving obstacle or predator, \Eq{dispersion} implies the construction shown in Fig.\ \ref{fig:conediagram}(b).  From this construction, we obtain the wedge angles, $\theta_{\pm}$, in a plane passing through the cone's axis (defined by $\bd{v}_0$ and $\bd{v}_p$), given by
\begin{equation}\label{eqn:thetapm}
\tan \theta_{\pm} = \pm \frac{1}{\cos (\psi)} \left( \frac{c_s}{v_r} \right) + \tan (\psi).
\end{equation}
Cross sections for the wedge shapes of both the static and moving obstacle are shown in Fig.\ \ref{fig:conediagram}.  In three dimensions, a moving obstacle produces an oblique circular cone as illustrated in Fig.\ \ref{fig:oblique}.  If $\bd{v}_0$ and $\bd{v}_p$ are co-linear, the the cone is a right-circular cone with $\theta_+ = \theta_{-}$, and $\psi = 0$.

\section{Linearized Theory in a Two Dimensional Flock}\label{sec:linear}
In order to assess the extent to which these predictions apply more generally, we consider a two-dimensional case for both the static and moving obstacle situations.  We first solve a linearized version of Eqs.\ (\ref{fullsystemnew1}) and (\ref{fullsystemnew2}) for the density fluctuation $\delta \rho/\rho_0$ in terms of an integral.  We then specialize to a static case to  analyze $\delta \rho / \rho_0$ analytically.  Following that, we numerically evaluate our integral-expression solution for both the static and moving cases and compare the results to our simple predictions above.

To specialize to two dimensions, we consider a flock equilibrium that is spatially uniform in the $z$ direction.  Suppose that an obstacle (call it a predator) is moving through the flock at constant velocity, $\bd{v}_p$, relative to the fixed frame of the medium (e.g.~air or water) in which the flock moves.  We assume that there is no motion in the $z$ direction.   We model the obstacle by a moving, localized, repulsive potential, $\eta(\bd{x},t)$, and add the term $-\nabla \eta$ to the right-hand side of \Eq{fullsystemnew1}.  We take $\eta$ to be Gaussian in space and given by
\begin{equation}
\eta = \eta (\bd{x},t) = \eta_0 \exp \left[ -\frac{(x-v_{p \, x} t)^2+(y - v_{p \, y} t)^2}{l^2} \right],
\end{equation}
where $\eta_0$ is the strength of the obstacle, $l$ is the length scale over which the obstacle acts, and $v_{p \, x}$ and $v_{p \, y}$ are the components of is the predator's velocity, $\bd{v}_p$, in the $x$-$y$ plane.  In two dimensions, this can roughly be thought of as a kind of moving flagpole around which the flock must navigate.  Without loss of generality, we set $\bd{v}_0  = v_0 \, \hat{\bd{x}}$.  Given this situation we consider the steady-state flock response in the approximation of infinite flock size.  The dynamics of a finite-size flock as it impinges on an obstacle hitting a flock has not been considered in the present work.  For a simulation of such a situation, see \cite{Lee06}. 

We add the obstacle potential to \Eq{fullsystemnew1} and, in the frame of the obstacle, linearize around the constant density, as we did in Sec.\ \ref{sec:dispersion}.  Taking a spatial Fourier transform of Eqs.\ (\ref{fullsystemnew1}) and (\ref{fullsystemnew2}) (including the obstacle), we obtain the following steady-state (i.e., $\partial/\partial t = 0$) equations, 
\begin{align}\label{FTedEqnsmovingObstacle}
\left( - i \bd{k} \cdot \bd{v}_p + i \bd{k} \cdot \bd{v}_0 \right) \, \delta \tilde{\bd{v}} &= -\frac{2}{\tau v_0^2} \left( \bd{v}_0 \cdot \delta \tilde{\bd{v}}\right) \bd{v}_0 - i \bd{k} \left( c_s^2 \frac{\delta \tilde{\rho}}{\rho_0} + \tilde{u} \delta \tilde{\rho} + \tilde{\eta} \right) - \nu_w \delta \tilde{\bd{v}}\\\label{deltarhomovingobstacle}
- i \bd{k} \cdot \bd{v}_p \, \delta \tilde{\rho} &+ i \bd{k} \cdot \bd{v}_0 \, \delta \tilde{\rho} + \rho_0 i \bd{k} \cdot \delta \tilde{\bd{v}} = 0,
\end{align}
with
\begin{equation}
\tilde{\eta}(k^2) = \eta_0 l^2 \pi e^{-\frac{1}{4} \left( k_x^2 + k_y^2 \right) l^2},
\end{equation}
and the other quantities defined in Eqs.\ (\ref{FTdefs} - \ref{FTdefsend}).  Using \Eq{deltarhomovingobstacle} to eliminate $\delta \tilde{\rho}/\rho_0$ from \Eq{FTedEqnsmovingObstacle}, we arrive at
\begin{equation}\label{mastermoving}
i \bd{k} \cdot \bd{v}_r \, \delta \tilde{\bd{v}} + \frac{2}{v_0^2 \tau} \left( \bd{v}_0 \cdot \delta \tilde{\bd{v}}\right) \bd{v}_0 + \nu_w \, \delta \tilde{\bd{v}} - i \bd{k} \left( c_s^2 + \rho \tilde{u} \right) \frac{\bd{k} \cdot \delta \bd{v}}{\bd{k} \cdot \bd{v}_r} = - i \bd{k} \tilde{\eta}.
\end{equation}
where, again, $\bd{v}_r = \bd{v}_0 - \bd{v}_p$.  Since there is no disturbance along the $z$-direction, we set $\bd{k} = k_x \hat{\bd{x}} + k_y \hat{\bd{y}}$.  Introducing an orthonormal basis \{$\hat{\bd{a}}_1$, $\hat{\bd{a}}_2$, $\hat{\bd{a}}_3$\} such that 
\begin{equation}
\hat{\bd{a}}_1 = \hat{\bd{k}} = \frac{\bd{k}}{k}, \quad \hat{\bd{a}}_2 \cdot \hat{\bd{z}} = 0, \quad \hat{\bd{a}}_3 = \hat{\bd{z}}, 
\end{equation}
we project \Eq{mastermoving} onto these three directions.  Defining
\begin{equation}
\delta \tilde{\bd{v}} = \delta \tilde{v}_1 \, \hat{\bd{a}}_1 + \delta \tilde{v}_2 \, \hat{\bd{a}}_2 + \delta \tilde{v}_3 \, \hat{\bd{a}}_3,
\end{equation}
\Eq{mastermoving} yields
\begin{align}
i k v_{r} \cos (\psi - \phi) \delta \tilde{v}_1 &+ \frac{2}{\tau v_0^2} v_{0} \cos (\phi) \left( v_{0} \cos (\phi) \delta \tilde{v}_1 - v_{0} \sin (\phi) \delta \tilde{v}_2 \right) \nonumber\\ \label{eqn:linearized1}
&+ \nu_w \delta \tilde{v}_1 - i k (c_s^2 + \rho_0 \tilde{u}) \frac{\delta \tilde{v}_1}{v_{r} \cos (\psi - \phi)} = -i k \tilde{\eta} \\\label{eqn:linearized2}
i k v_{r} \cos (\psi - \phi) \delta \tilde{v}_2 &- \frac{2}{\tau v_0^2} v_{0} \sin (\phi) \left( v_{0} \cos (\phi) \delta \tilde{v}_1 - v_{0} \sin (\phi) \delta \tilde{v}_2 \right) + \nu_w \delta \tilde{v}_2 = 0 
\end{align}
and $\delta \tilde{v}_3 = 0$, where we have changed coordinates from $(k_x,k_y)$ to $(k, \phi)$.  Here $\phi$ is the angular orientation of $\bd{k}$, measured from the $x$ axis, and $\psi$ is the angle between $\bd{v}_r$ and the $x$ axis.  We can obtain general results for $\delta \bd{v} (x,y)$ and $\delta \rho(x,y)$ by solving Eqs.\ (\ref{eqn:linearized1}), (\ref{eqn:linearized2}), and (\ref{deltarhomovingobstacle}) for $\delta \tilde{v}_1$, $\delta \tilde{v}_2$, and $\delta \tilde{\rho}$ and then inverse Fourier transforming the result.  However, for simplicity in what follows, we again take $\tau \rightarrow 0$, clamping all of the flocking individuals to the same speed.  Equations (\ref{eqn:linearized1}), (\ref{eqn:linearized2}), and (\ref{deltarhomovingobstacle}) then yield
\begin{equation}\label{eqn:integrand}
\frac{\delta \tilde{\rho} (k,\phi)}{\rho_0} = \frac{k^2 \tilde{\eta} \sin^2 (\phi)}{k^2 v_r^2 \cos^2 (\psi - \phi)-k^2 \bar{c}^2 \sin^2 (\phi) - i k \nu_w v_r \cos (\psi - \phi)},
\end{equation}
with 
\begin{equation}\label{cbar}
\bar{c}^2 (k) = c_s^2+\rho_0 \tilde{u} (k) = c_s^2 + \frac{4 \pi \rho_0 u_0}{k^2 + \kappa_{\rho}^2}.
\end{equation}
By inverse Fourier transforming, we obtain the density perturbation at any point $(r,\theta)$ in the flock,
\begin{equation}\label{mainintegral}
\frac{\delta \rho (r,\theta)}{\rho_0} = \frac{1}{(2 \pi)^2} \int_0^{\infty} \!\!\! \int_0^{2 \pi} \frac{\delta \tilde{\rho} (k,\phi)}{\rho_0} \, e^{i k r \cos (\phi - \theta)} k d\phi d k,
\end{equation}
where $\delta \tilde{\rho} (k,\phi) /\rho_0$ is defined in \Eq{eqn:integrand}.  In the next section, we explore \Eq{mainintegral} analytically in the case of a static obstacle.  After that, we evaluate \Eq{mainintegral} numerically for both a static and a moving obstacle/predator.

\section{Analytical Results for a Static Obstacle}
To evaluate the integral, we consider the following illustrative case.  We take $\bd{v}_r = \bd{v}_0$, which corresponds to a stationary obstacle or predator.  This implies that $\psi = 0$.  Also, assume that the parameters are such that for most values of $k$, the quantities $\bar{c}$ and $\nu_{w}$ can be approximated by their large $k$ limits.  We have
\begin{align}
\bar{c} &\approx \lim_{k \rightarrow \infty} \bar{c} (k) = c_s,\\
\nu_{w} &\approx \lim_{k \rightarrow \infty} \nu_{w} (k) = \frac{4 \pi w_0}{\kappa_w^2}.
\end{align}
The range over which this approximation is good will be investigated in Sec.\ \ref{sec:num}.  With these approximations we can integrate (\ref{mainintegral}) to obtain the density perturbation $\delta \rho / \rho_0$.  
\begin{figure}[!hb]
	\begin{center}
 	\includegraphics[width=12cm]{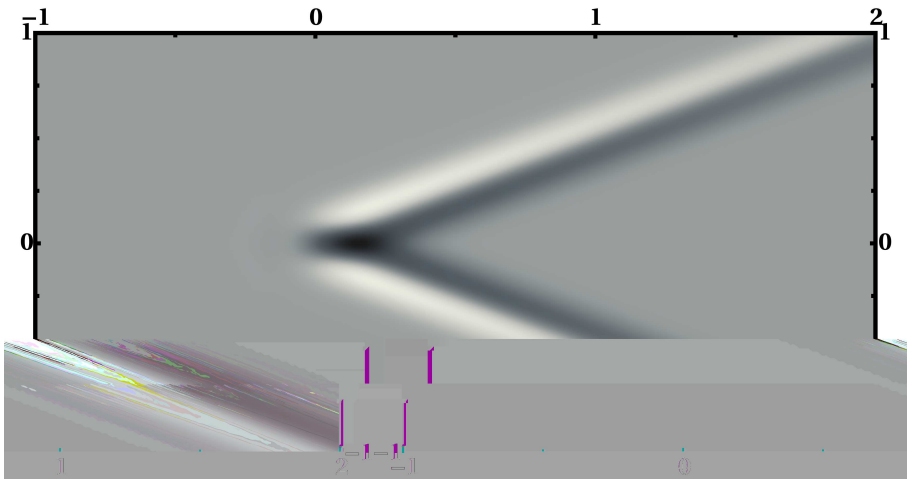}\\
      \caption{\label{fig:analytic}Plot of the density fluctuations.  Parameters used for the figure are: $c_s = 15$, $\eta_0 = 1$, $\epsilon = 0.838$, $\gamma = 2$, and $l=0.1$.  The $x$ axis is horizontal, the $y$ axis is vertical.}
	\end{center}
\end{figure}
The full analysis is done in the Appendix.  Fig.\ \ref{fig:analytic} displays the density perturbation for a particular choice of the parameters.  Note that the main feature is a wedge formed from the information of the obstacle propagating through the flock.  We find, in the Appendix, that density will be large when $y$ is near $\pm y_0$, where 
\begin{equation}\label{eqn:y0}
y_0 (x) = \frac{1}{\gamma} \left( x - \frac{l^2}{2 \gamma} \epsilon \right).
\end{equation}
where we have introduced the quantities $\epsilon = \nu_{w} /2 \bar{c}$ and $\gamma = v_0 /\bar{c}$.  The first term in \Eq{eqn:y0}, $x/\gamma$, corresponds to the wedge condition, \Eq{tantheta}.

\begin{figure}[!hb]
	\begin{center}
 	\includegraphics[width=12cm]{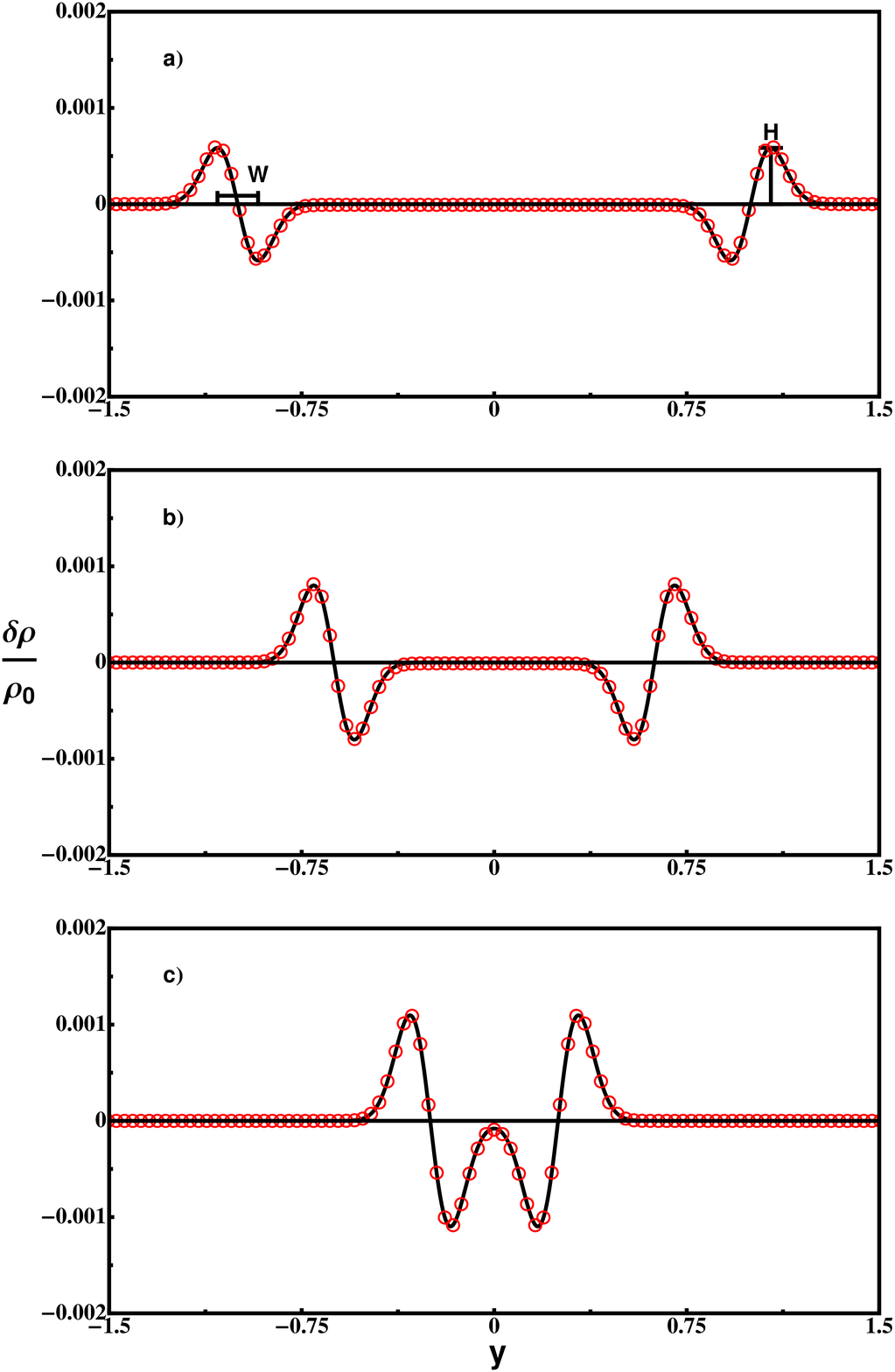}\\
      \caption{\label{fig:analyticprofiles}(color online) Plot of the density fluctuation for constant values of $x$ for the same parameters in Fig.\ \ref{fig:analytic}.  The red circles are numerical values (computed via Sec.\ \ref{sec:num} methods) and the solid curve is the theory. a) $x=2$.  The definitions for height and width are displayed on the plot. b) shows $x=1.25$, and c) $x=0.5$.}
	\end{center}
\end{figure}

Figure \ref{fig:analyticprofiles} shows $\delta \rho(x,y)/\rho_0$ versus $y$ for several fixed values of $x$.  Numerical data (computed in Sec.\ \ref{sec:num}) are plotted as open circles, and the theory obtained in the Appendix is plotted as a solid curve.  They agree well.  Further approximations (see the Appendix) result in analytic expressions for the height, $\mathcal{H}$, and width, $\mathcal{W}$, of these profiles (illustrated in Fig.\ \ref{fig:analyticprofiles}(a)).  
The height, $\mathcal{H}$ ($\delta \rho/\rho_0$ at maximum), and width, $\mathcal{W}$ (distance between maximum and minimum), are
\begin{align}\label{eqn:height}
\mathcal{H}(x) &= \frac{\eta_0 \gamma}{\bar{c}^2 (1+\gamma^2)} \sqrt{\frac{\pi}{2}} \exp \left[ -\frac{1}{2} + \frac{\epsilon^2 l^2-4 x \gamma \epsilon}{4 \gamma^2} \right] \\\label{eqn:width}
\mathcal{W} &= \frac{2}{\sqrt{2}} \frac{\sqrt{1+\gamma^2}}{\gamma} l.
\end{align}

From these expressions we see that the width is predicted to be insensitive to many of the parameters of our problem except $l$ and $\gamma$.  For example, the width does not increase far from the source of the disturbance (i.e., $W$ in \Eq{eqn:width} does not depend on $x$).  A main feature of \Eq{eqn:height} is its prediction of the exponential decay of the height of the disturbance with increasing $x$.  In the next section, we compare numerical simulations with these predictions.

\section{Numerical Results}\label{sec:num}
\begin{figure}[!hb]
\begin{center}
 	\includegraphics[width=12cm]{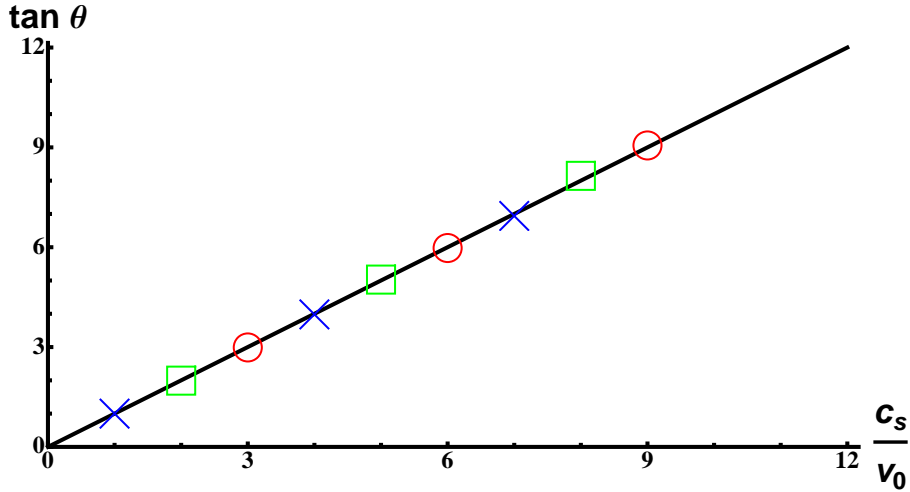}\\
	\caption{\label{fig:staticwedgeangle}(color online) Plot showing the wedge angle $\theta$ vs.~the quantity $\frac{c_s}{v_0}$, for $c_s = 15$.  This shows the agreement of the numerical data for $w_0 = 0.001$ (red circles), $w_0 = 0.01$ (green boxes), and $w_0 = 0.1$ (blue crosses).  The solid line is the theoretical prediction for the wedge angle from \Eq{eqn:thetapm}.  The other parameters for the plot are $u_0 = 0.1$, $\kappa_{\rho} = 0.1$, and $\kappa_w = 0.5$, $l=0.1$, $\eta_0 = 1$, and $\rho_0 = 0.8$.}
\end{center}
\end{figure}
\begin{figure}[!hb]
	\begin{center}
 	\includegraphics[width=12cm]{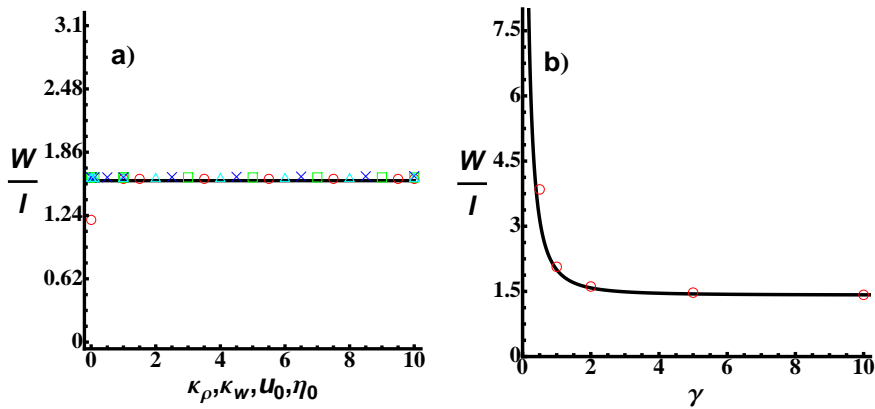}\\
      \caption{\label{fig:w}(color online) Graphs showing the dependence of the width on various parameters of the model.  The solid curves is the expression in \Eq{eqn:width}.  The colored markers are numerical values obtained using the processes described in Sec.\ \ref{sec:num}.  a) $\mathcal{W}/l$ vs.~$\kappa_\rho$ (squares), $\kappa_{w}$ (circles), $u_0$ (crosses), and $\eta_0$ (triangles). b) $\mathcal{W}/l$ vs.~$\gamma$. If a parameter is not varied then it has the value: $\rho_0 = 0.8$, $c_s = 15$, $\gamma = 2$, $\eta_0 = 1$, $u_0 = 0.001$, $w_0 = 0.5$, $\kappa_{w} = 0.5$, $\kappa_{\rho} = 0.1$, $l=0.1$, and $x=2$.}
	\end{center}
\end{figure}
\begin{figure}[!hb]
	\begin{center}
 	\includegraphics[width=10cm]{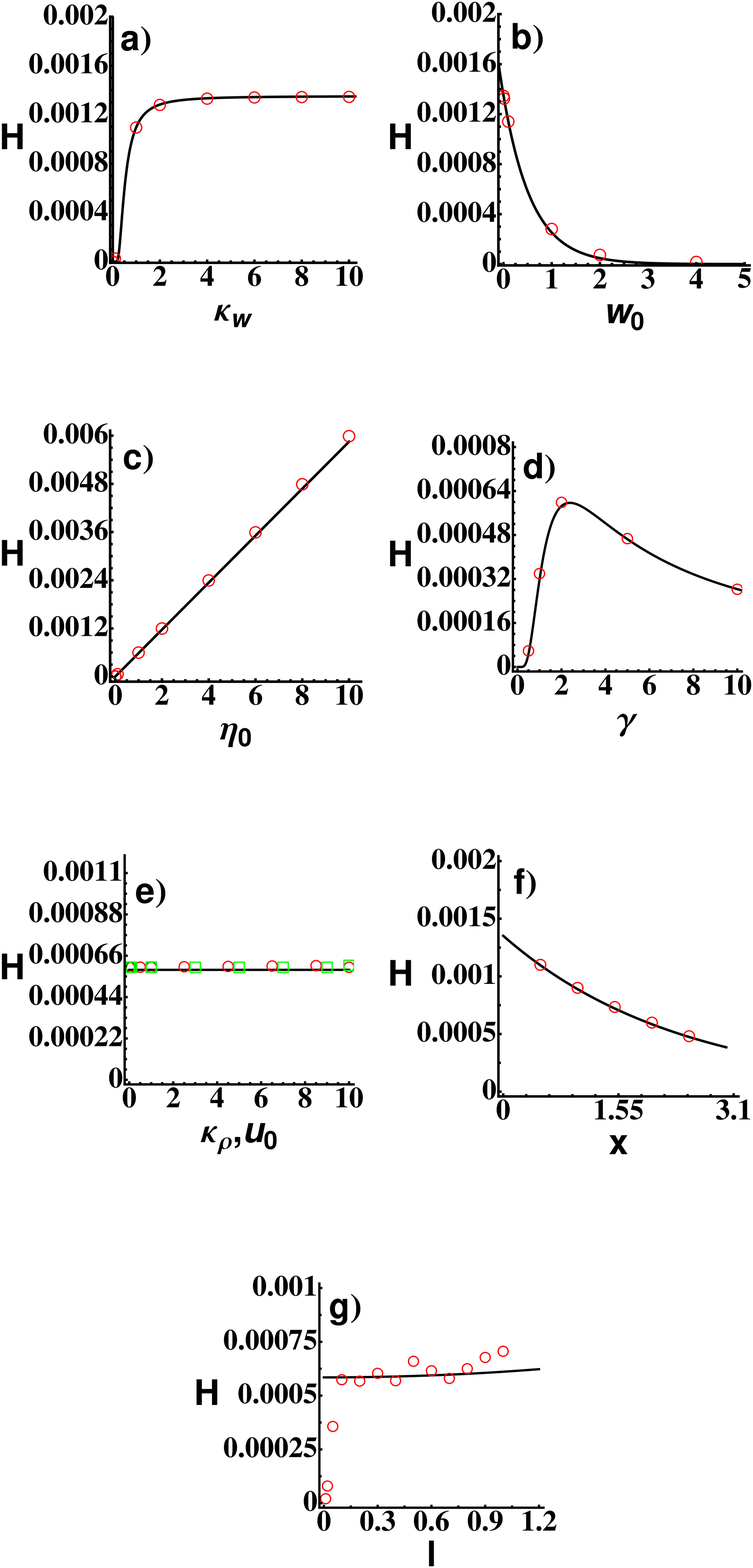}\\
      \caption{\label{fig:h}(color online) Various graphs showing the dependence of the height of the wedge on various parameters of the model.  The solid curves is the expressions in \Eq{eqn:height}.  The colored markers are numerical values obtained using the processes in Sec.\ \ref{sec:num}.  a) $\mathcal{H}$ vs.~$\kappa_w$. b) $\mathcal{H}$ vs.~$w_0$. c) $\mathcal{H}$ vs.~$\eta_0$. d) $\mathcal{H}$ vs.~$\gamma$. e) $\mathcal{H}$ vs.~$\kappa_\rho$ (boxes), and $u_0$ (circles). f) $\mathcal{H}$ vs. $x$. g) $\mathcal{H}$ vs.~$l$.  If a parameter is not varied then it has the value: $\rho_0 = 0.8$, $c_s = 15$, $\gamma = 2$, $\eta_0 = 1$, $u_0 = 0.001$, $w_0 = 0.5$, $\kappa_{w} = 0.5$, $\kappa_{\rho} = 0.1$, $l=0.5$, and $x=2$.}
	\end{center}
\end{figure}

\subsection{The Static Obstacle}\label{numerics}
In order to evaluate the integral in \Eq{mainintegral} numerically, we need to do a two-dimensional infinite integral at each point in physical space.  To do this, we express the kernel of the inverse Fourier transform as a sum of Bessel functions using the Jacobi-Anger expansion (see, for example, \cite{abramowitz+stegun}).  This allows us to do one of the iterated integrals via contour integration.  We then obtain an infinite sum of single integrals at each real space point that we evaluate numerically.  An example of the density fluctuation evaluated using these methods looks very similar to Fig.\ \ref{fig:analytic}.  Similar to the analytic result in the Appendix, the numerical density fluctuation shows a prominent wedge emanating from near the origin.  The correspondence to the analytic work is excellent and can be seen in Fig.\ \ref{fig:analyticprofiles}.  We now compare the numerical results to theoretical predictions for $\theta_{\pm}$, $\mathcal{H}$, and $\mathcal{W}$.

Using the relation in \Eq{tantheta}, we can test the above results to determine the accuracy of the numerical fit to the wedge angle predicted earlier.  Visually determining the angle from the numerical output gives a well defined wedge angle to about $0.5^\circ$ accuracy.  The tangent of this angle, \Eq{tantheta}, can then be compared with the quantity $c_s/v_0$.  Figure \ref{fig:staticwedgeangle} shows that this comparison yields very good agreement for the static case.  The parameters used in Fig.\ \ref{fig:staticwedgeangle} are $w_0 = 0.1$ (blue crosses), $w_0 = 0.01$ (green boxes), and $w_0 = 0.001$ (red circles).  It is seen that changing $w_0$ leaves the wedge angle essentially unchanged. 

Figure \ref{fig:w} shows comparisons between results for $\mathcal{W}$ from the numerical simulations (colored markers) with the predictions given in \Eq{eqn:width} (solid curves).  Figure \ref{fig:w}(a) and shows that, as predicted by the theory, $\mathcal{W}/l$ is insensitive to the values of $\kappa_{\rho}$, $\kappa_{w}$, $u_0$, and $\eta_0$.  The only important dependence of the width was on the parameter $\gamma$ as seen in Fig.\ \ref{fig:w}(b).  Here we see that the wedge width approaches a constant value for large $\gamma$.  Figure \ref{fig:h}(a-d) show the dependence of the height, $\mathcal{H}$, on $\kappa_w$, $w_0$, $\eta_0$, and $\gamma$, respectively.  In these figures, as well as in Fig.\ \ref{fig:w}, if a parameter is not varied, then it has the value: $\rho_0 = 0.8$, $c_s = 15$, $\gamma = 2$, $\eta_0 = 1$, $u_0 = 0.001$, $w_0 = 0.5$, $\kappa_{w} = 0.5$, $\kappa_{\rho} = 0.1$, $l=0.1$, and $x=2$.  As predicted by \Eq{eqn:height}, $\mathcal{H}$ is linear in $\eta_0$.  Figure \ref{fig:h}(e) shows that the height is insensitive to both $\kappa_\rho$ and $u_0$.  The theoretical prediction for the dependence of the height on position, $x$, is verified in Fig.\ \ref{fig:h}(f).  Figure \ref{fig:h}(g) shows that there is agreement with \Eq{eqn:height} for $l \gtrsim 0.15$, but breaks down at small $l$ since the width is predicted to go to zero in that case.  Additionally, the expansion in the Appendix used to obtain \Eq{resultformula} implies that our approximations are expected to become invalid at large $\epsilon/\gamma =  2 \pi w_0 / \kappa_{w}^2 v_0$.  For example, at very low $\kappa_w$, the width starts deviating from the prediction (Fig.\ \ref{fig:w}(a), circles).

\subsection{The Moving Obstacle}
\begin{figure}[!ht]
\begin{center}
	\includegraphics[width=12cm]{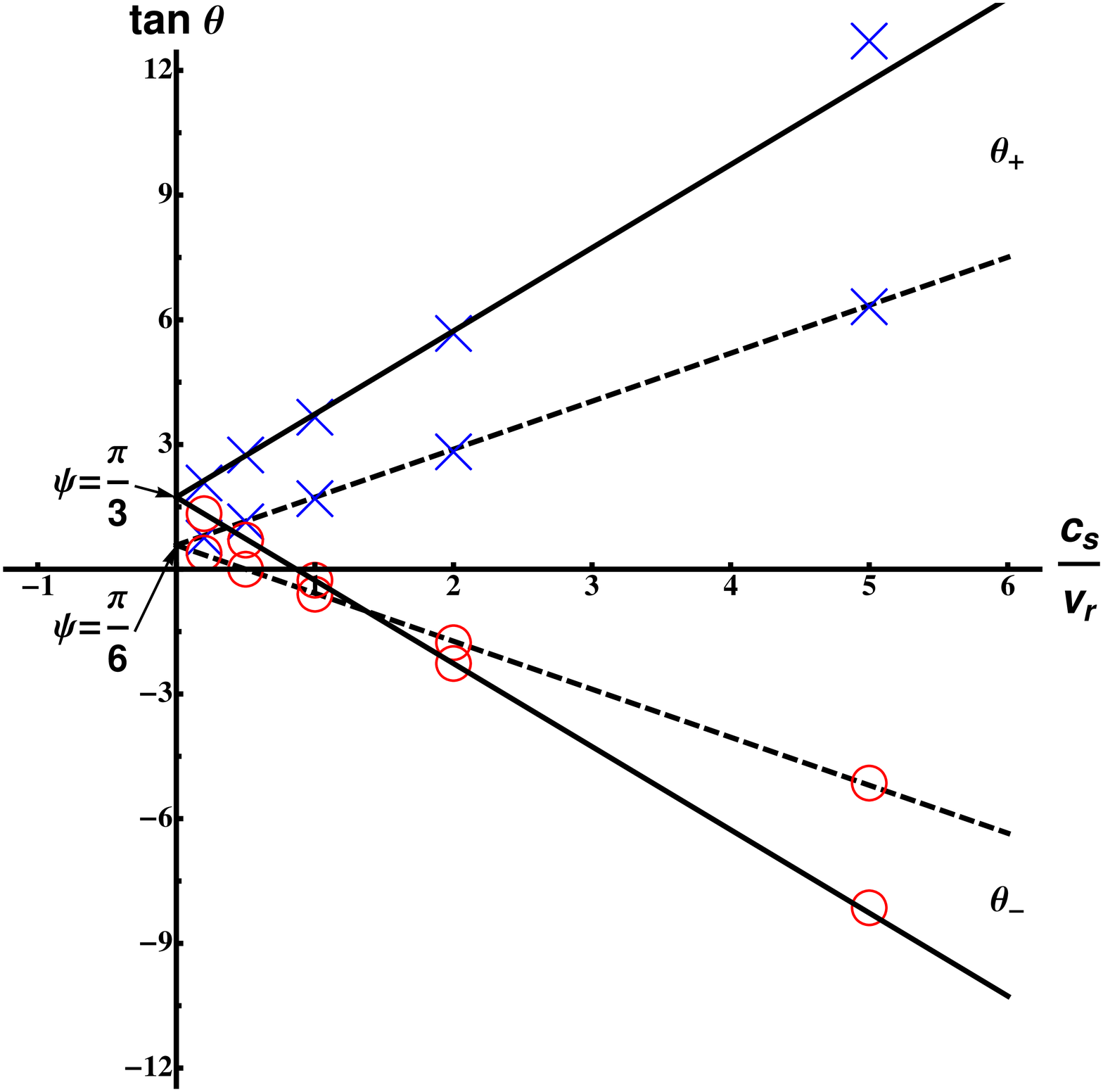}\\
	\caption{\label{fig:wedgeanglemoving}(color online) Plot showing the wedge angles $\theta_{\pm}$ vs the quantity $\frac{c_s}{v_r}$.  This shows the agreement of the numerical data for two choices of predator angle $\psi$.  The solid lines are the prediction of \Eq{eqn:thetapm} for $\psi= \pi/3$, whereas the dashed lines are for $\psi = \pi/6$.  The lines with the positive slope correspond to $\theta_{+}$ (blue crosses), and the lines with the negative slope correspond to $\theta_{-}$ (red circles).  The lines are the theoretical prediction for the wedge angles.  The other parameters for these plots were $w_0 = 0.001$, $u_0 = 0.1$, $\kappa_\rho = 0.1$, and $\kappa_w = 0.1$.}
\end{center}
\end{figure}

We numerically evaluated \Eq{mainintegral} using the same method as Sec.\ \ref{numerics}, but with nonzero predator angle, $\psi$.  The results of the comparison between the theoretical prediction of the wedge angles and the numerics can be seen in Fig.\ \ref{fig:wedgeanglemoving}.  The theory, \Eq{eqn:thetapm}, predicts the wedge angles as a function of predator angle, $\psi$.  The figure shows the correspondence to the numerical data for two angles, $\psi = \pi/3$ and $\psi = \pi/6$, versus various values of $c_s/v_0$.  The agreement is excellent, and, similar to the static case, the wedge angles are insensitive to parameters such as the nonlinear viscosity parameters, $w_0$ and $\kappa_w$.

\section{Conclusions}
In this paper we have explored the response of a flock to static and moving obstacles.  The obstacle is introduced into a flock and the fluctuations about an equilibrium are analyzed.  We find that with both the static and the moving obstacles, the flock produces a prominent wedge where the information is propagating away from the disturbance, as shown by Fig.\ \ref{fig:analytic}.  The wedge angles can be predicted using a simple geometric construction.  The information/disturbance propagates asymmetrically (unless $\psi = 0$), with two angles, $\theta_+$ and $\theta_-$, given by \Eq{eqn:thetapm}.  We tested this analytically as well as numerically, and the result is found to agree well with the theoretical prediction.  The wedge angles are insensitive to most physical parameters, most notably the velocity viscosity term, $\bd{W}$, and, unlike the well-known Mach cone in acoustics, there is no threshold speed for existence.  Specifically, the angles only depend on the speed of sound in the flock, $c_s$, the speed of the flock, $v_0$, the relative speed of the obstacle to the flock, $v_r$, and the angle between them $\psi$.  Heights and widths of the Mach cones for $\psi = 0$ are given by the analytic expression in \Eq{eqn:height} and \Eq{eqn:width}.  Numerical results are in good agreement with these expressions.  It is also noteworthy that the wedge width, defined in Fig.\ \ref{fig:analyticprofiles}(a), is insensitive to many parameters in the model as can be seen in Fig.\ \ref{fig:w}(a).

Future work should include the dynamics of an obstacle hitting a flock, extension to $\tau \neq 0$, and a physical explanation of the wedge shape and offset from the origin.  Finally, the extension to a fully nonlinear treatment of the obstacle is of interest.

This work was financially supported by ONR grant N00014-07-1-0734.


\appendix
\section{Derivation of $\delta \rho/\rho_0$, the Height, and the Width of the Disturbance}
\begin{figure}[!hb]
	\begin{center}
 	\includegraphics[width=12cm]{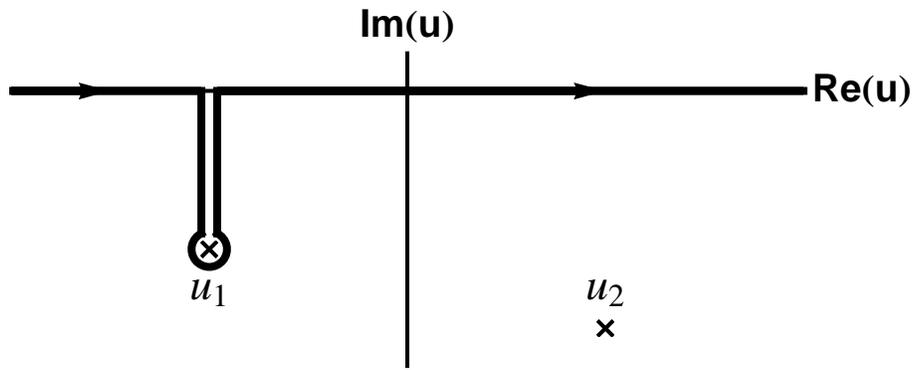}\\
      \caption{\label{fig:contour}Contour in complex plane for the integral in \Eq{eqn:contour}.  This contour is forced by causality and from $y \rightarrow -y$ symmetry.}
	\end{center}
\end{figure}

We derive the density perturbation $\delta \rho / \rho_0$, the height, $\mathcal{H}$, and the width, $\mathcal{W}$, via a direct computation of the integral \Eq{mainintegral}.  To evaluate the integral, we consider the following special case.  First, we take $\bd{v}_r = \bd{v}_0$.  This implies that $\psi = 0$.  Also, as in the main body of the paper,
\begin{align}
\bar{c} &\approx \lim_{k \rightarrow \infty} \bar{c} (k) = c_s,\\
\nu_{w} &\approx \lim_{k \rightarrow \infty} \nu_{w} (k) = \frac{4 \pi w_0}{\kappa_w^2},
\end{align}
and we define $\epsilon = \frac{\nu_{w}}{2 \bar{c}}$ and $\gamma = \frac{v_0}{\bar{c}}$.  With these approximations we can write the integral (\ref{mainintegral}) in rectangular coordinates,
\begin{equation}\label{cartintegral}
\frac{\delta \rho(r,\theta)}{\rho_0} = -\frac{\eta_0 \pi l^2}{\bar{c}^2 \left( 2 \pi \right)^2} \int_{-\infty}^{\infty} e^{i k_x x} e^{-\frac{l^2}{4} k_x^2} \int_{-\infty}^{\infty} \frac{k_y^2 e^{i k_y y-\frac{l^2}{4}k_y^2}}{k_y^2 - k_x^2 \gamma^2 + \frac{i \nu_{w} \gamma}{\bar{c}} k_x} d k_y d k_x.
\end{equation}
Let us do the $k_y$ integral first.  We shift the path of integration up in the complex $k_y$ plane to $\textrm{Im}(k_y) = i\frac{2 y}{l^2}$ so as to go through the saddle point in the complex plane giving,
\begin{equation}\label{cartintegral2}
\frac{\delta \rho(r,\theta)}{\rho_0} = -\frac{\eta_0 l^2}{4 \bar{c}^2 \pi} \int_{-\infty}^{\infty} e^{i k_x x} e^{-\frac{l^2}{4} k_x^2} \left( \int_{-\infty}^{\infty}  \frac{\left(u + i \frac{2 y}{l^2} \right)^2 e^{-\frac{l^2}{4} u^2}}{\left( u - \hat{u}_1 \right) \left( u - \hat{u}_2 \right)} d u \right) d k_x,
\end{equation}
where the integral is over real $u$, we have factored the denominator, and we define
\begin{align}\label{eqn:radical1}
\hat{u}_1 &= -\gamma k_x \sqrt{1-\frac{i 2 \epsilon}{\gamma k_x}} - \frac{i 2 y}{l^2},\\\label{eqn:radical2}
\hat{u}_2 &= \gamma k_x \sqrt{1-\frac{i 2 \epsilon}{\gamma k_x}} - \frac{i 2 y}{l^2}.
\end{align}
Using
\begin{equation}\label{eqn:contour}
\int \frac{(u+i a)^2 e^{- b^2 u^2}}{(u - u_1)(u - u_2)} du = -\frac{i \pi}{u_1 - u_2} \left[ (a - i u_1)^2 w (b u_1) + (a - i u_2)^2 w(-b u_2) \right] + \frac{\sqrt{\pi}}{b}
\end{equation}
along the contour given in Fig.\ \ref{fig:contour}, we can explicitly evaluate the $u$ integral in terms of the complex error function $w$, given by (see \cite{abramowitz+stegun})
\begin{equation}
w(z) = \frac{i}{\pi} \int_{-\infty}^{\infty} \frac{e^{-t^2}}{z-t} \, dt = e^{-z^2} \textrm{erfc} (- i z) \quad \textrm{if $\textrm{Im} (z) > 0$}
\end{equation}
and defined for the negative imaginary $z$ by analytic continuation.  

Expanding \Eq{eqn:radical1} and \Eq{eqn:radical2} as Taylor series in $(\epsilon / \gamma k_x)$, we obtain from \Eq{cartintegral2}
\begin{equation}\label{resultformula}
\frac{\delta \rho }{\rho_0} = -\frac{\eta}{\bar{c}^2} e^{-\frac{(x^2 + y^2)}{l^2}} + A(x) e^{-B \left[ y-y_0(x) \right]^2} \left( \int_{-\infty}^{\infty} e^{-w^2} f_{+} dw \right) + A(x) e^{-B \left[ y+y_0(x)\right]^2} \left( \int_{-\infty}^{\infty} e^{-w^2} f_{-} dw \right)
\end{equation}
where 
\begin{equation}
f_{\pm}(w) = C_{\pm} (w) \textrm{erfc} (F_{\pm} (w)),
\end{equation}
and 
\begin{align}
y_0 &= \frac{1}{\gamma} \left( x - \frac{l^2}{2 \gamma} \epsilon \right) \\
A(x) &= \frac{\eta_0 l}{4 \bar{c}^2 \sqrt{1+ \gamma^2}} \exp \left[ -\frac{\epsilon}{\gamma} (x-\frac{l^2}{4 \gamma} \epsilon) \right] \\
B &= \frac{\gamma^2}{(1+\gamma^2) l^2} \\
C_{\pm} (w) &= i \frac{2 \gamma}{l \sqrt{1 + \gamma^2}} w - \frac{\gamma (x - \frac{l^2}{2 \gamma} \epsilon)}{\frac{l^2}{2} (1 + \gamma^2)} \pm \frac{\gamma^2 y}{\frac{l^2}{2} (1 + \gamma^2)} \\
F_{\pm} (w) &= i \frac{\gamma}{\sqrt{1+\gamma^2}} w - \frac{\gamma (x - \frac{l^2}{2 \gamma} \epsilon)}{l (1 + \gamma^2)} \mp \frac{y}{l (1 + \gamma^2)}.
\end{align}

We now approximate the integral over $w$ using, for example, integration formula 25.4.46 on pg.~890 of \cite{abramowitz+stegun}, to obtain an analytic expression for $\delta \rho / \rho_0$.  All non-numerical plots and images are composed via this method (using $n=10$).  We can further use
\begin{equation}
\textrm{erf} (u+i v) \approx \textrm{erf} (u) + \frac{e^{-u^2}}{2 \pi u} \left[ (1-\cos (2 u v)) + i \sin (2 u v) \right] \approx \textrm{erf} (u)
\end{equation}
from pg.~299, of the same text, to approximate the integrand.  For large enough $x$, if we change variables and shift the origin in the $y$ direction to the center of the wedge, $y_0$, we see that the real part of the argument of the error function is
\begin{equation}
-\frac{y_0 (x)}{l} - \frac{\bar{y}}{l (1 + \gamma^2)},
\end{equation}
where $y = y_0 + \bar{y}$.  Since for modest values of $x$ this is typically far from zero, the error function is approximately constant and equal to 2.  This gives (near the center of the wedge for fixed $x$)

\begin{equation}
A(x) e^{-B \bar{y}^2} \int_{-\infty}^{\infty} e^{-w^2} f_{+}(w) dw \approx A(x) e^{-B \bar{y}^2} \frac{4 \gamma^2 \sqrt{\pi}}{l^2 (1+\gamma^2)} (\bar{y})
\end{equation}
where we have integrated a Gaussian, and neglected the imaginary part, since the final integral must be real.  Thus, the height, $\mathcal{H}$ ($\delta \rho/\rho_0$ at maximum), and width, $\mathcal{W}$ (distance between maximum and minimum) defined in Fig.\ \ref{fig:analyticprofiles}, are given in the main body of the paper (\Eq{eqn:height} and \Eq{eqn:width}) as
\begin{align*}
\mathcal{H}(x) &= \frac{\eta_0 \gamma}{\bar{c}^2 (1+\gamma^2)} \sqrt{\frac{\pi}{2}} \exp \left[ -\frac{1}{2} + \frac{\epsilon^2 l^2-4 x \gamma \epsilon}{4 \gamma^2} \right] \\
\mathcal{W} &= \frac{2}{\sqrt{2}} \frac{\sqrt{1+\gamma^2}}{\gamma}l.
\end{align*}
\end{document}